# A STUDY OF THE GAIN OF MICROCHANNEL PLATES IN THE IONIZATION PROFILE MONITORS AT FERMILAB*

R. Thurman-Keup†, D. Slimmer, C. Lundberg, J.R. Zagel
Fermi National Accelerator Laboratory, Batavia, USA

*Abstract*

One of the on-going issues with the use of microchannel plates (MCP) in the ionization profile monitors (IPM) at Fermilab is the significant decrease in gain over time. There are several possible issues that can cause this. Historically, the assumption has been that this is aging, where the secondary emission yield (SEY) of the pore surface changes after some amount of extracted charge. Recent literature searches have brought to light the possibility that this is an initial 'scrubbing' effect whereby adsorbed gasses are removed from the MCP pores by the removal of charge from the MCP. This paper discusses the results of studies conducted on the IPMs in the Main Injector at Fermilab.

## INTRODUCTION

Ionization profile monitors (IPM) are used in many accelerator laboratories around the world [1-7]. They have been used in nearly all the synchrotrons built at Fermilab, and presently are used in the Main Injector (MI), Recycler Ring (RR), and Booster synchrotron [8], with another one being planned for the Integral Optics Test Accelerator (IOTA). All Fermilab IPMs, as well as many of those at other laboratories, utilize one or more microchannel plates (MCP) for signal amplification. Historically we have found that the gain of the MCP decreases over time (Fig. 1) and have attributed it to the well-known fact that they age with current extracted from them [9]. Thus, we have periodically replaced them.

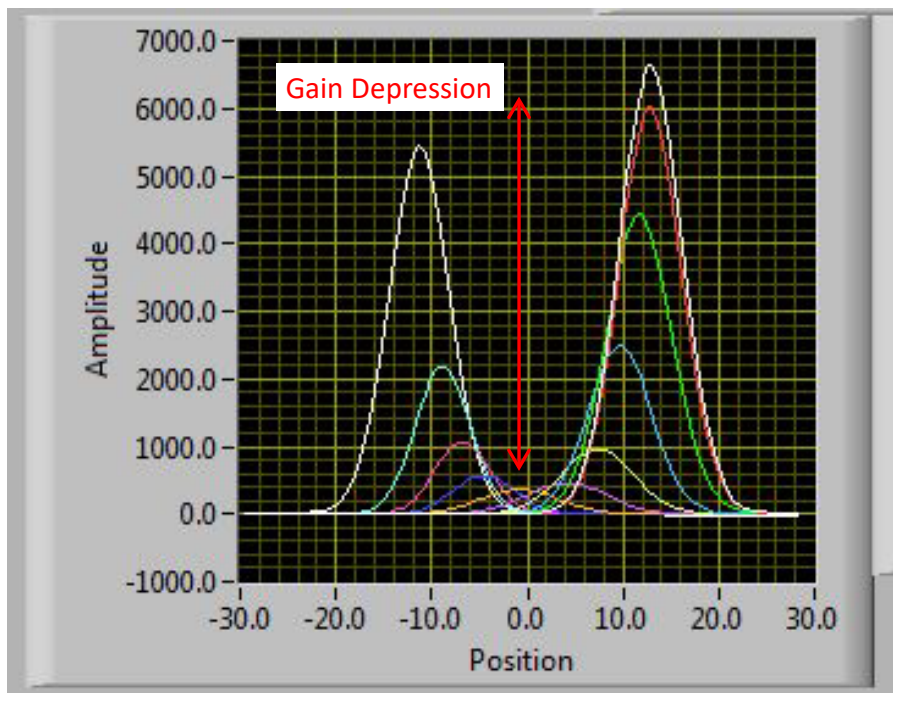

Figure 1: Scan of MCP plate across beam showing gain depression where beam is normally positioned.

Recently, more in-depth investigations have revealed that the decrease in the gain is much more consistent with conditioning, or 'scrubbing', of the MCPs, and not aging. Literature searches have rediscovered the fact that a decrease in gain with conditioning is a known property of MCPs [10,11]. Our own historical IPM data and a recent dedicated test show results which are much more consistent with what one expects from conditioning. In this paper, we summarize our current understanding of the behavior of the MCPs in our IPMs.

## EXPERIMENTAL DEVICE

IPMs are devices which utilize the ionization of residual gas to measure the transverse profile of the beam. Figure 2 is a diagram of the present Fermilab IPMs. When the beam passes through the IPM, it ionizes the residual gas. The IPM collects the ionization products using an electrostatic field to accelerate them to the MCP.

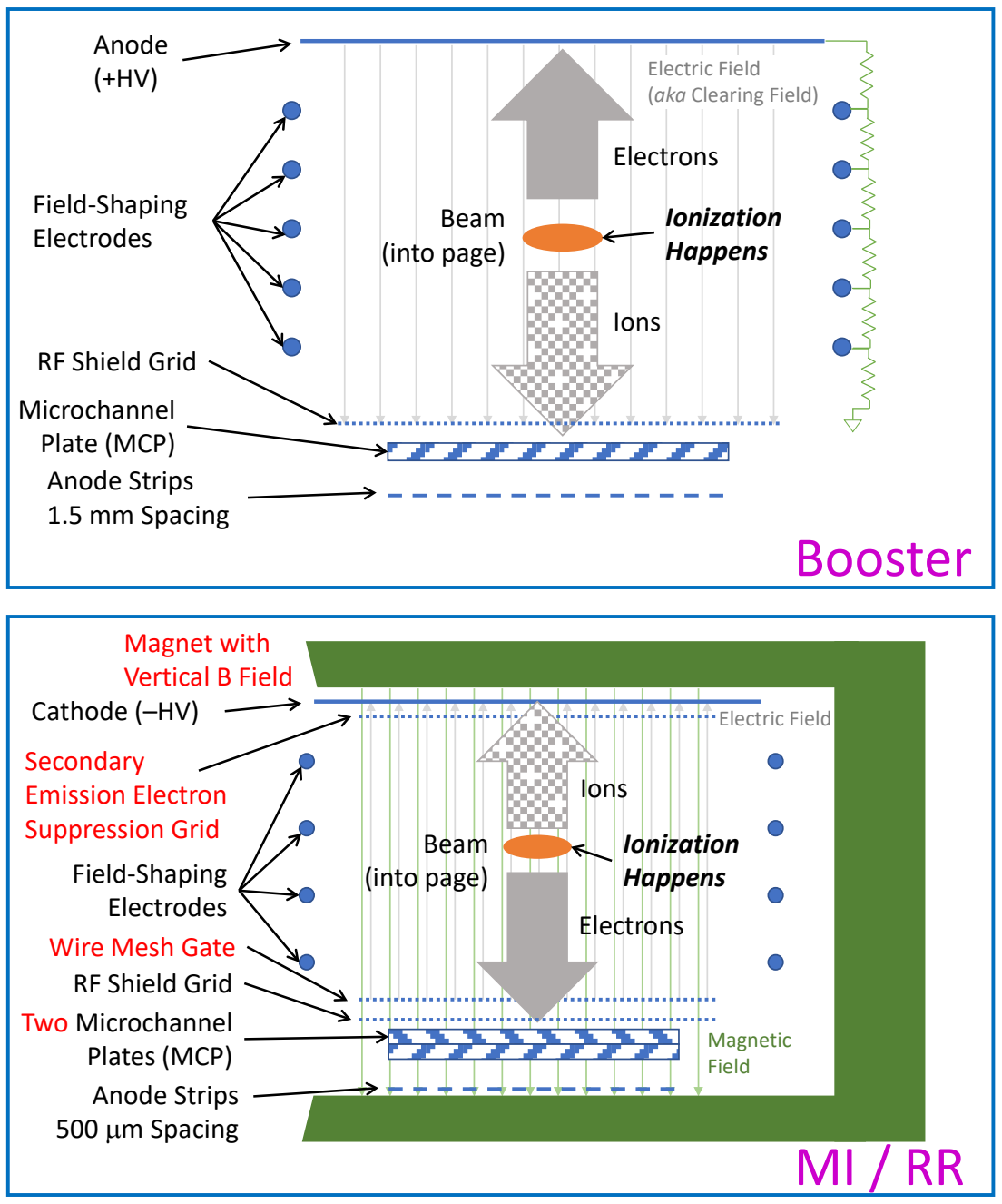

Figure 2: Diagrams of Booster and MI/RR IPMs.

The MCP is a thin plate with microscopic holes (aka pores) through it, each of which acts as a charge multiplier, much like a photomultiplier tube (PMT) (Fig. 3) [12,13].

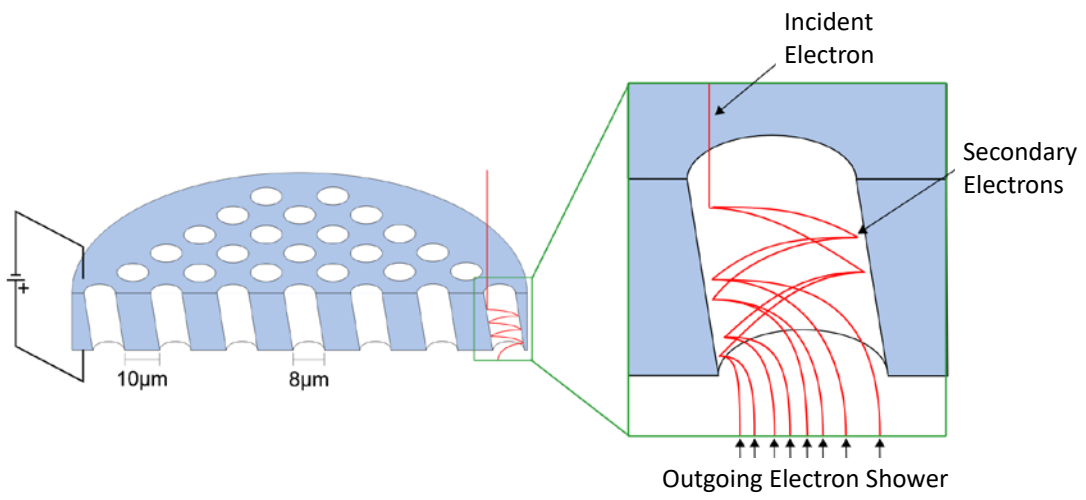

Figure 3: Schematic of MCP functional behavior showing amplification by electron avalanche.

† keup@fnal.gov

The output electrons from the MCP are collected via thin conductive strips which function as the anode in the multiplication chain. This current is converted to a voltage by a preamp and digitized. The digitized signals from the IPM are processed by a LabVIEW program that generates gaussian profile fits to the data. The IPMs collect data on request. The high voltage is turned on and data is collected for a single injection-extraction cycle which is usually less than one second.

There are two IPM versions at Fermilab (as shown in Fig. 2): magnetic and electrostatic. The magnetic version, used in MI and RR, collects electrons, and has both an electric field to accelerate the electrons and a magnetic field to constrain their trajectories. Without the magnetic field, the electrons would be moved transversely by the beam space charge and not maintain the position where the ionization occurred. These IPMs are also installed on motorized stages such that the MCP can be moved relative to the beam. The electrostatic version, used in Booster, collects ions, and has no magnetic field since the ions are heavy and not dramatically affected by the beam space charge. These are bolted to the beampipe and cannot be moved relative to the beam.

## PRIOR MCP LITERATURE

MCPs have been used since at least the early 1970's for image intensification [14]. A mathematical model of MCPs was developed in reference [11]. In addition to the model, the paper also showed the measured gain of an MCP at various stages in the preparation: initial, after vacuum bake, after scrub, and after final seal-in. The gain decrease from after vacuum bake to after scrub was a factor of 3. Scrubbing refers to the initial phase of operation of the MCP, where the electrons act to ionize adsorbed gases and remove them from the surfaces of the pores. Since these gases are additional contributors of secondary emission electrons, the gain is higher with them than without them. One approach to this conditioning is to illuminate the MCP with ultraviolet radiation until the gain stabilizes [10]. Whether or not the MCP can be exposed to atmosphere after conditioning without losing the benefits of the conditioning is not yet entirely clear to us [15,16]. Ideally the scrubbing would happen after installation of the MCP in vacuum. The use of MCPs in satellites encounters similar gain issues as we do with the IPMs [17-19]. When used in a spectrometer, spectral lines are placed at defined positions on the MCP. As it is apparently impractical [16] to keep a satellite MCP under vacuum, the initial operation of it in space causes the brighter lines to scrub faster resulting in a non-uniform gain, just as the beam in an accelerator is always located at the same location and produces a dip in the gain. Reference [20] discusses gain issues impacting MCP-based PMTs used in the PANDA experiment at FAIR. They also consider the new atomic layer deposition technique for MCPs. The lifetime of a MCP is generally stated in units of C/cm$^2$ of extracted charge and is usually understood to be > 1 C/cm$^2$. Conditioning of the MCP requires much less than that, typically much less than 0.1 C/cm$^2$ [9,19,21].

## IPM DATA

The present versions of the IPMs were installed in the MI and RR in 2014, and the Booster in 2017. All the data collected since then is available for analysis and is what is used to assess the behavior of the MCPs.

### *Historical Booster IPM Data*

The MCPs in the Booster IPMs were not changed until 2022, providing 5 years of MCP gain data. Two things simplify the analysis of Booster IPM data: each IPM has only one MCP, and the Booster operation is always the same. Figure 4 shows a typical IPM signal for a Booster cycle.

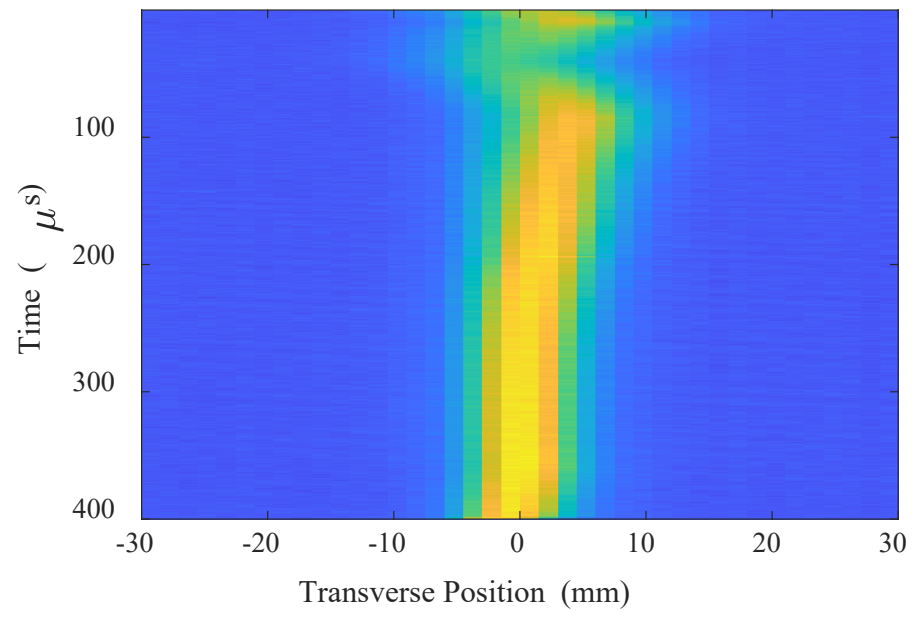

Figure 4: Booster cycle from injection (top) to extraction (bottom).

The goal of the analysis is to measure the relative gain of the MCP over time, ideally as a function of extracted charge. To determine the gain, one must scale the signal to the beam intensity, and adjust for the voltage setting of the MCP. The beam intensity is recorded for each acquisition. In addition, there were occasional measurements made at a series of voltage settings, which are used to determine the gain vs. voltage curve (Figs. 5 and 6).

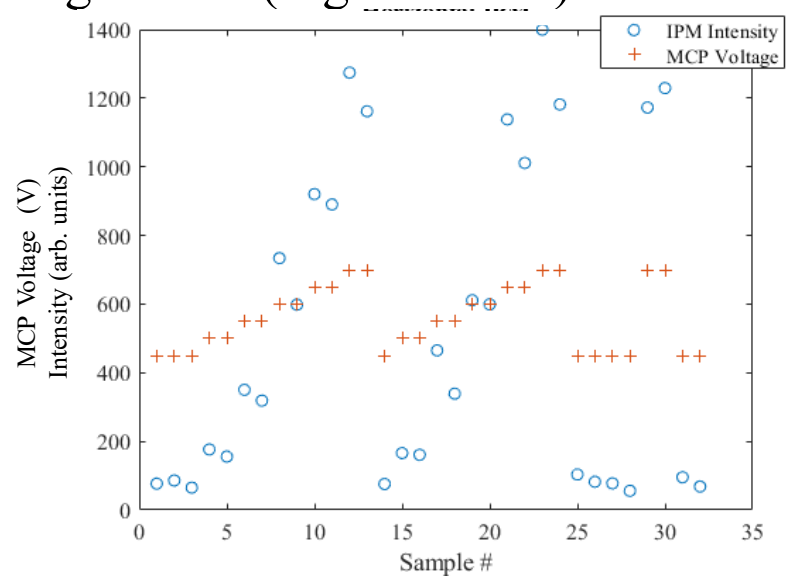

Figure 5: Collection of 32 acquisitions with varied MCP voltage. From this we extract a voltage gain curve.

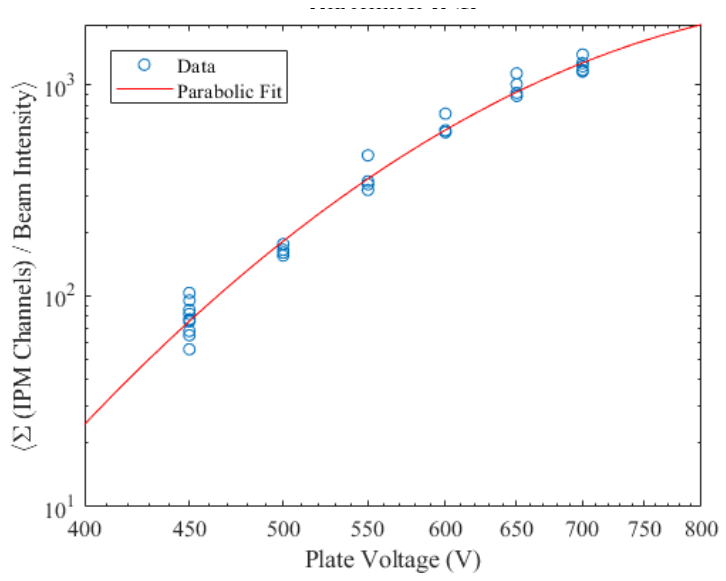

Figure 6: MCP voltage vs. gain curve with parabolic fit. This fit is used to normalize the IPM data to extract the MCP gain.

Figure 7 shows the distribution of IPM acquisitions as a function of time. We calculated the MCP gain from this data by scaling it by the beam intensity and by the voltage gain of Fig. 6. The resulting MCP gain is shown in Fig. 8 as a function of time.

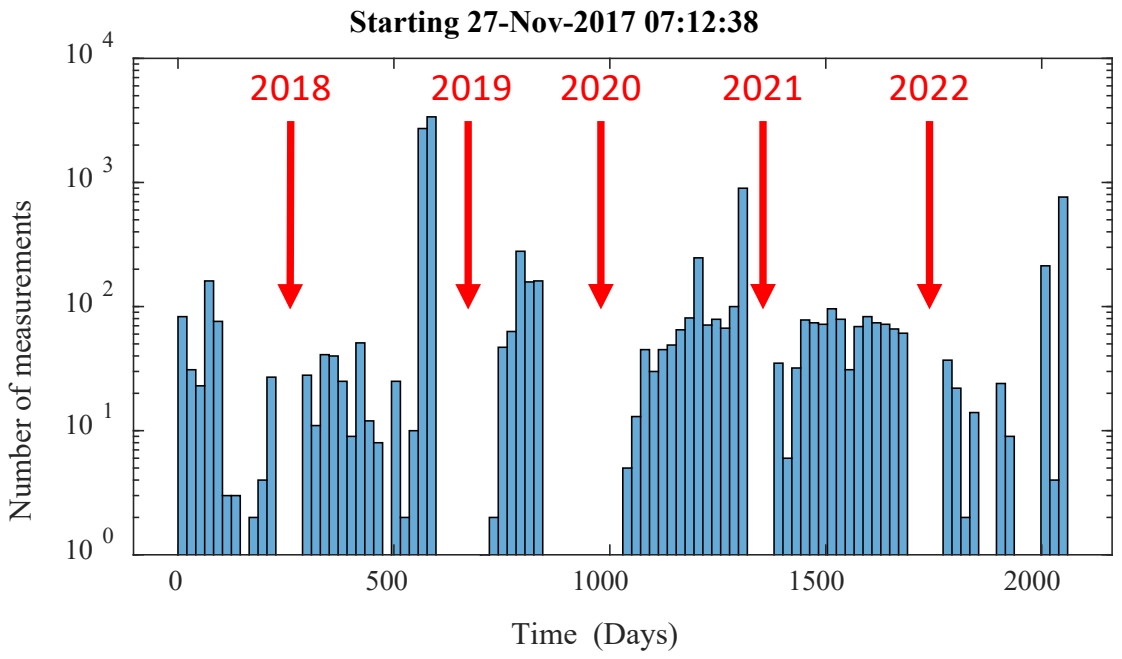

Figure 7: The red arrows indicated maintenance periods which are usually July – October.

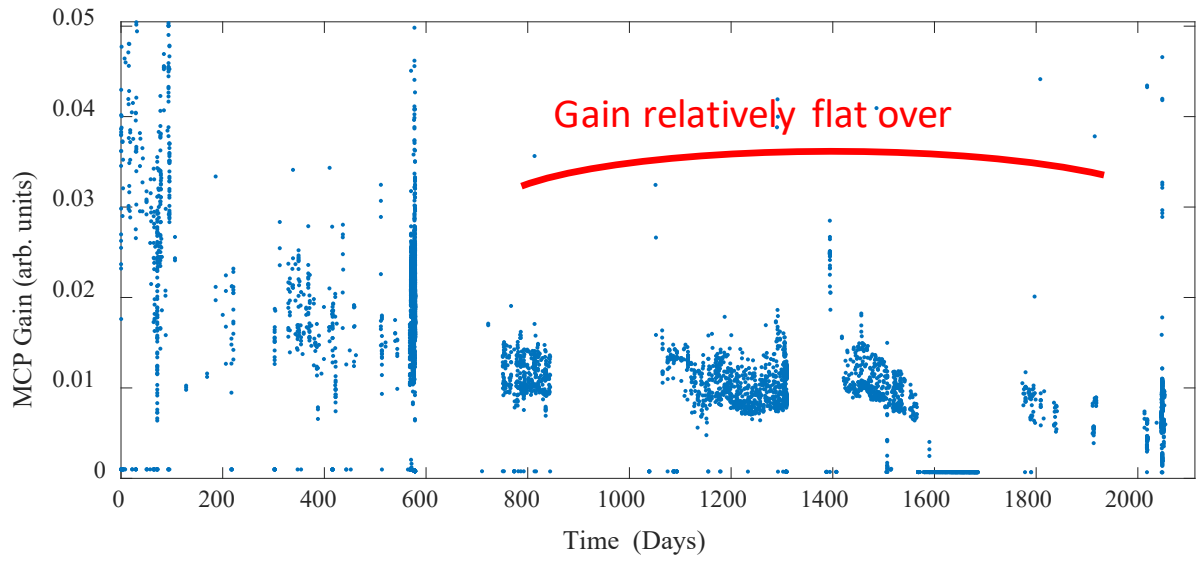

Figure 8: Relative MCP gain as a function of time. The gain has been relatively flat for the last 4 years.

From this plot, it would appear that the MCP gain decreased after initial installation but has been relatively flat after that. This is a behavior consistent with initial conditioning, and not aging. One thing that the reader might notice is the slight decrease over the last 600 days or so. It was discovered late in the writing of this paper that scaling the IPM signal by the measured beam intensity is not entirely accurate. As one can see in the Fig. 9, there is a clear non-linear relationship even after scaling, and, in addition, there are isolated regions, all of which must still be investigated. The non-linear relationship is in part responsible for the decrease in the last 200 days where the beam intensity was much less for most of the acquisitions. Despite this behavior, we don't see a dramatic decrease in the apparent gain of the MCP over recent time.

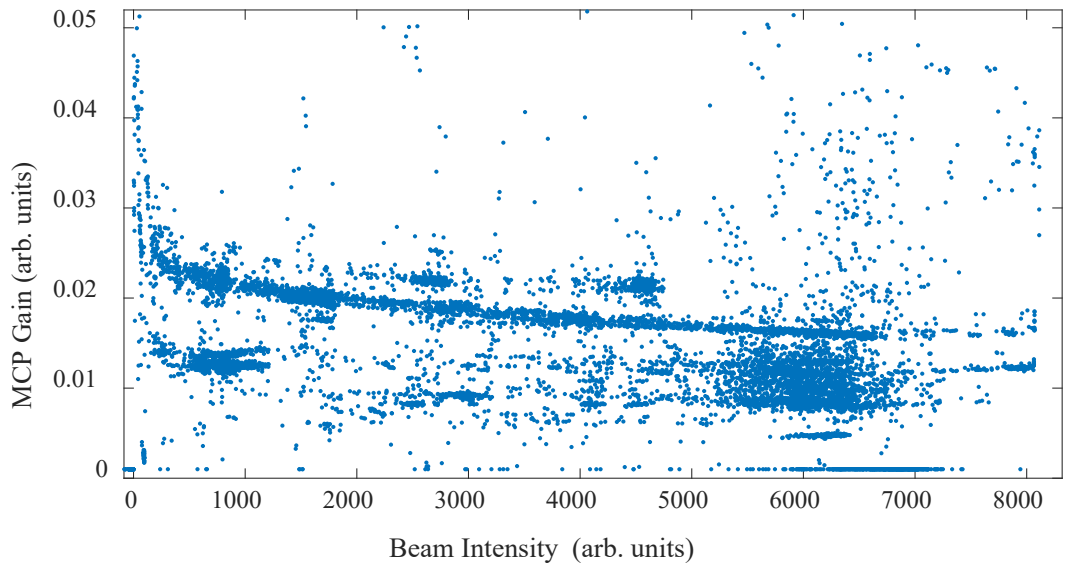

Figure 9: Relationship between MCP response and beam intensity after correcting the MCP response for the beam intensity and voltage.

Since the MCP should age with extracted charge, we calculate the extracted charge for an acquisition as $Q = ITD$, where $I$ is the average current from the MCP during the acquisition, $T$ is the time the high voltage is on (~3 seconds), and $D$ is the Booster duty cycle (~50%). Figure 10 shows the integral of the extracted charge vs. time. At present, the total extracted charge is a little over 4 mC/cm$^2$, which is not yet near the level where aging would be a concern.

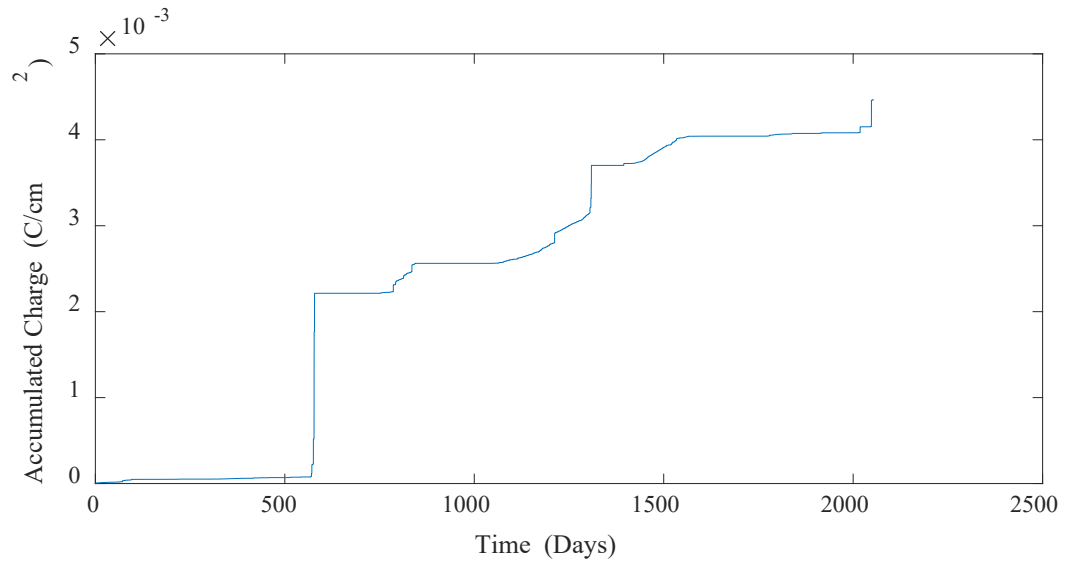

Figure 10: Integral of the extracted charge as a function of time.

## *MI Dedicated Experiment*

Recently, one of the MI IPMs was run continuously to measure the behavior of the gain. Since these IPMs can be moved, the MCP was moved to a new region of the plate, and then run repeatedly for a period of ~5 days. Figure 11 shows the typical MI IPM signals where one can see that the MCP has been positioned to the side of the beam.

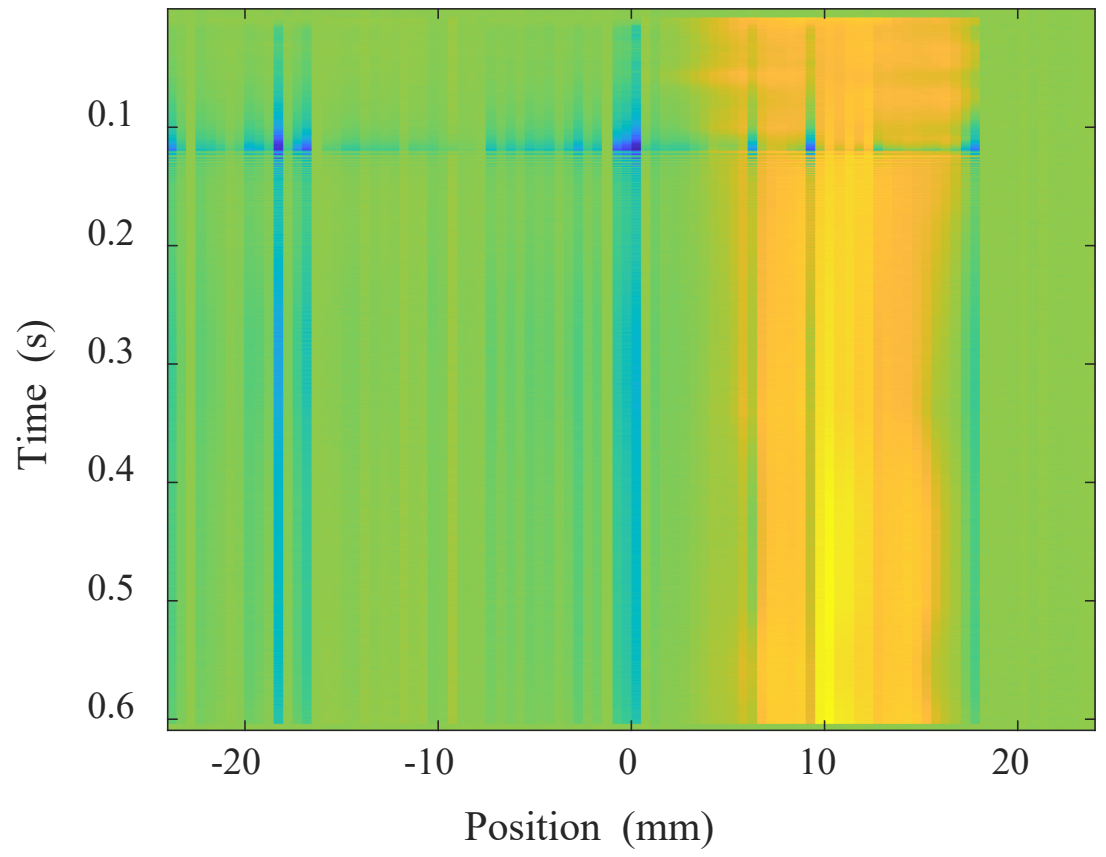

Figure 11: MI cycle from injection (top) to extraction (bottom). Some channels have obvious problems (blue).

Figure 12 shows the raw IPM integrated signals, the beam intensities, and the processed signals (i.e. the MCP gain). During the 5 days, the voltage was periodically adjusted to keep the signals at similar levels. The processing corrected for the beam intensity and accounted for the changes in voltage by scaling the data in each voltage region to match at the boundaries. Figure 13 shows this gain as a function of acquisition number, time, and integrated charge out of the MCP. Here as well, the change in gain is consistent with scrubbing, not with aging. One additional thing to note is that since the MI/RR IPMs have a pair of MCPs, the scrubbing is mostly affecting the second MCP which has much higher current draw. As such the gain will continue to decrease with time albeit much more slowly.

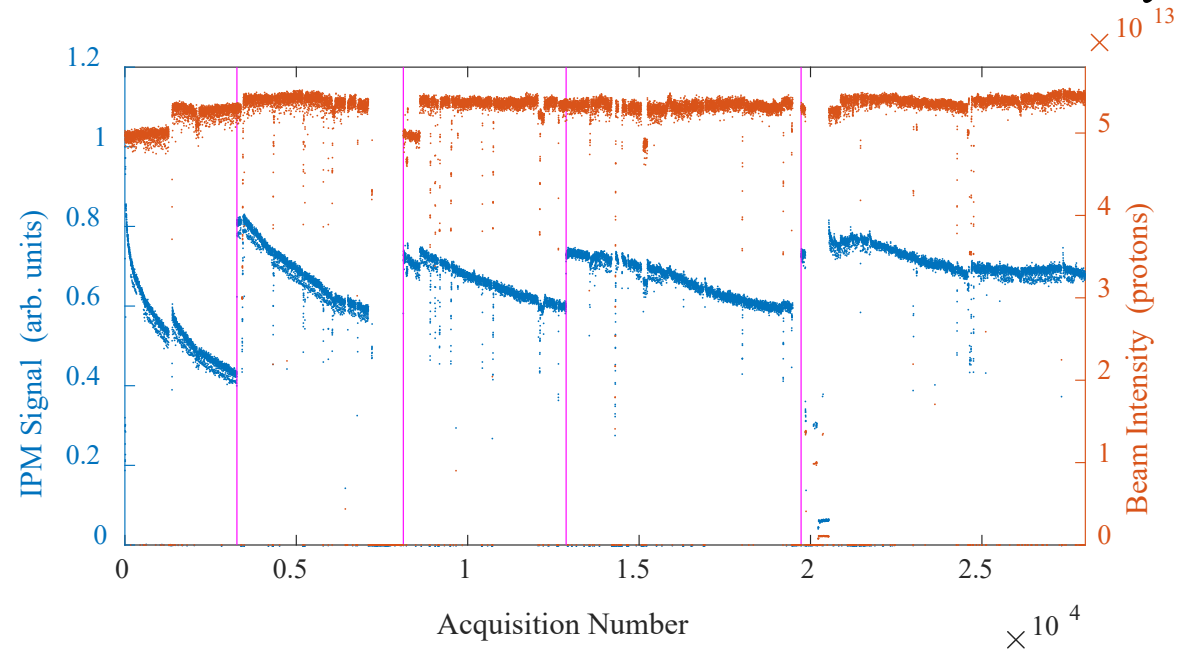

Figure 12: MI IPM signal (blue) and beam intensity (red). The magenta vertical lines indicate when the MCP voltage was changed.

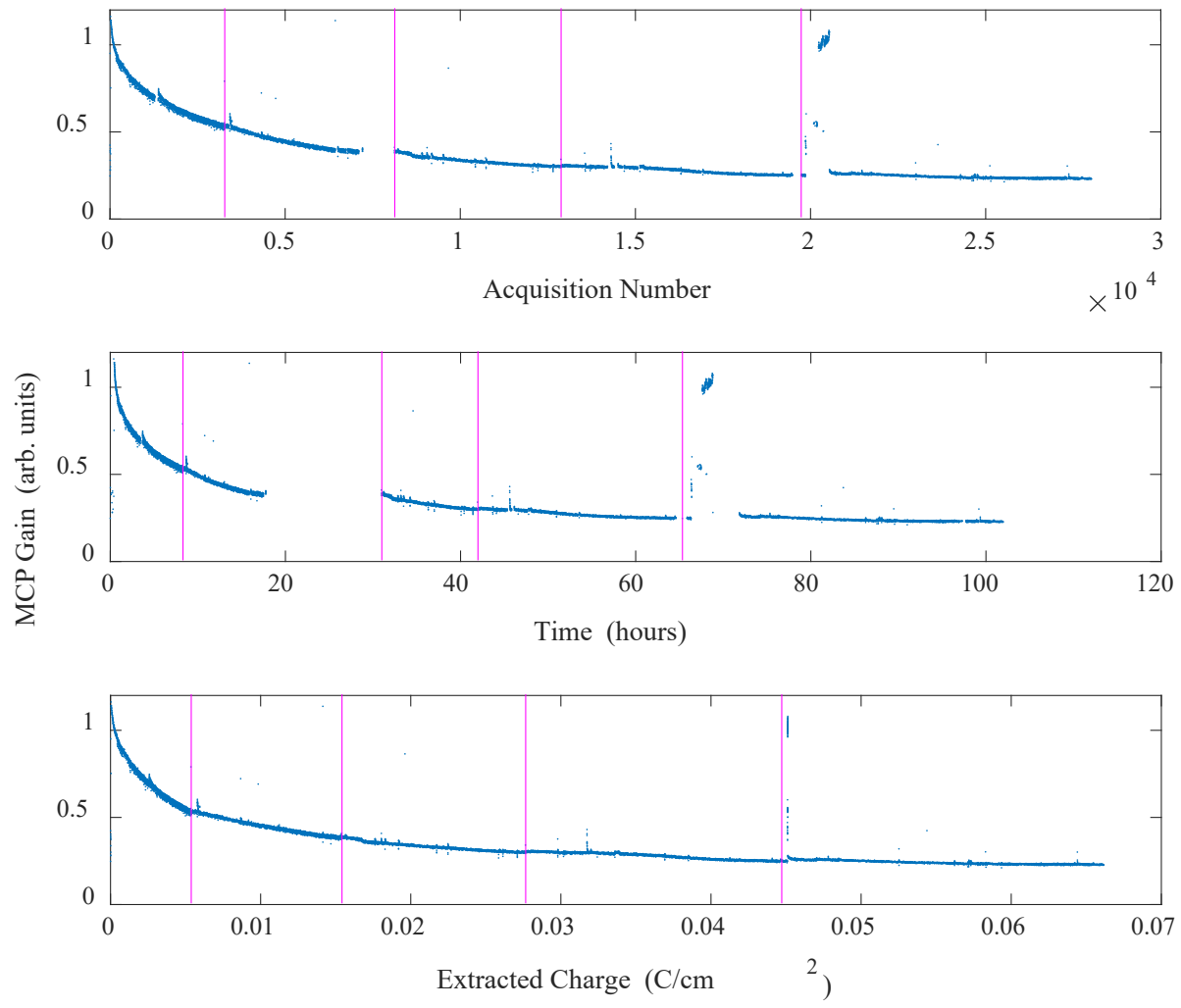

Figure 13: Integral of the extracted charge as a function of time.

## CONCLUSIONS

Both previous literature and the studies done with the IPMs indicate that aging is not the cause of the decrease in gain seen in our IPMs. The implication is that the MCPs do not need to be replaced as often as they have been in the past. Instead, our attention now focuses on the problem of *in situ* conditioning of the MCP or periodic calibration of the gain of the MCP.

As stated earlier, the MI and RR MCP devices are mounted on motorized stages and therefore can be moved. They can be slowly moved across the beam to allow the scrubbing to uniformly condition the MCP and avoid uneven gains. The same technique could also be used to calibrate the gain as a function of position, which may be a more useful approach since the gain will continue to decrease with use.

In the Booster, there is no way to easily change the relative position of the beam on the MCP other than to move the beam itself. However, if one lowers the electrostatic field, the ions will spread out transversely due to space charge and more uniformly cover the MCP. This may offer a way to initially scrub the MCP.

## REFERENCES

[1] K. Satou, "Development of a Gated IPM System for J-PARC MR", in *Proc. 8th Int. Beam Instrumentation Conf. (IBIC'19)*, Malmö, Sweden, Sep. 2019, pp. 343-346. doi:10.18429/JACoW-IBIC2019-TUPP020

[2] H.S. Sandberg *et al.*, "Commissioning of Timepix3 Based Beam Gas Ionisation Profile Monitors for the CERN Proton Synchrotron", in *Proc. 10th Int. Beam Instrumentation Conf. (IBIC'21)*, Pohang, Rep. of Korea, May 2021, pp. 172-175. doi:10.18429/JACoW-IBIC2021-TUOA05

[3] A. Jansson *et al.*, "IPM Measurements in the Tevatron", in *Proc. 22nd Particle Accelerator Conf. (PAC'07)*, Albuquerque, NM, USA, June 2007, pp. 3883-3885. doi:10.1109/PAC.2007.4440026

[4] M. Sachwitz, A.Hofmann, S. Pauliuk, K. Tiedtke and H. Wabnitz, "Ionization Profile Monitor to Determine Spatial and Angular Stability of FEL Radiation of Flash", in *Proc. 11th European Particle Accelerator Conf. (EPAC'08)*, Genoa, Italy, June 2008, pp. 1266-1268.

[5] R. Connolly, J. Fite, S. Jao, S. Tepikian and C. Trabocchi, "Residual-Gas-Ionization Beam Profile Monitors in RHIC", in *Proc. 14th Beam Instrumentation Workshop (BIW'10)*, Santa Fe, NM, USA, May 2010, pp. 116-118.

[6] K. Wittenburg, "Experience with the Residual Gas Ionisation Beam Profile Monitors at the DESY Proton Accelerators", in *Proc. 3rd European Particle Accelerator Conf. (EPAC'92)*, Berlin, Germany, Mar. 1992, pp.1133-1135.

[7] F. Benedetti *et al.*, "Design and Development of Ionization Profile Monitor for the Cryogenic Sections of the ESS Linac", *EPJ Web Conf.*, vol. 225, p. 01009, 2020. doi:10.1051/epjconf/202022501009

[8] J.R. Zagel *et al.*, "Third Generation Residual Gas Ionization Profile Monitors at Fermilab", in *Proc. 3rd International Beam Instrumentation Conference (IBIC'14)*, Monterey, CA, USA, Sep. 2014, paper TUPD04, pp. 408–411. https://jacow.org/IBIC2014/papers/tupd04.pdf

[9] B.R. Sandel, A. Lyle Broadfoot and D. E. Shemansky, "Microchannel Plate Life Tests", May 1977 / Vol. 16, No. 5 / APPLIED OPTICS, pp. 1435-1437.

[10] O.H.W. Siegmund, "Preconditioning of Microchannel Plate Stacks", Proc. SPIE 1072, Image Intensification, (April 1989), pp. 111-118. doi:10.1117/12.952545

[11] E.H. Eberhardt, "Gain Model for Microchannel Plates", APPLIED OPTICS / Vol. 18, No. 9 / 1 May 1979, pp. 1418-1423.

[12] J.L. Wiza, "Microchannel Plate Detectors", Nuclear Instruments and Methods, Vol. 162, 1979, pages 587 to 601.

[13] T. Gys, "Micro-channel Plates and Vacuum Detectors", Nuclear Instruments and Methods in Phyics Research A787(2015), pp.254–260. doi:10.1016/j.nima.2014.12.044

[14] J. M. Grant, in Proceedings of NASA ERDL Image Intensifier Symposium, 24-26 October 1961 (NASA Office of Scientific and Technical Information, Wash., D.C.), p. 63.

[15] J.P. Rager and J.F. Renaud, "The Use of a Microchannel Electron Multiplier in Spectroscopic Instrumentation, Involving Frequent Vacuum Breaking", Rev. Sci. Instrum., Vol. 45, No.7, July 1974, pp. 922-926.

[16] A. Chumikov, V. Cheptsov and N. Managadze, “Microchannel Plate Detector Gain Decrease Through Storage Under Environmental Conditions”, IEEE TRANSACTIONS ON INSTRUMENTATION AND MEASUREMENT, VOL. 72, 2023, p. 7003208. doi: 10.1109/TIM.2023.3265633

[17] N. Griffiths, S. Airieau and O. Siegmund, “In-flight performance of the SUMER microchannel plate detectors”, Proc. SPIE 3445, EUV, X-Ray, and Gamma-Ray Instrumentation for Astronomy IX, (10 November 1998), pp. 566-577. doi: 10.1117/12.330312

[18] J. De Keyser *et al.*, “Position-Dependent Microchannel Plate Gain Correction in Rosetta's ROSINA/DFMS Mass Spectrometer”, International Journal of Mass Spectrometry 446 (2019) 116232. doi:10.1016/j.ijms.2019.116232

[19] R.F. Malina and K.R. Coburn, “Comparative Lifetesting Results for Microchannel Plates in Windowless EUV Photon Detectors”, IEEE Transactions on Nuclear Science, Vol. NS-31, No. 1, February 1984, pp. 404-407.

[20] A. Lehmann *et al.*, “Recent Developments with Microchannel-plate PMTs”, Nuclear Instruments and Methods in Physics Research A 876 (2017) pp. 42–47. doi:10.1016/j.nima.2016.12.06

[21] G.W. Fraser, J.F. Pearson and J.E. Lees, “Evaluation of Long Life ($L^2$) Microchannel Plates for X-Ray Photon Counting”, IEEE Transactions on Nuclear Science, Vol. 35, No. 1, February 1988, pp. 529-533.